\documentclass[usenatbib]{mn2e}
\usepackage{amsmath}
\usepackage{graphicx}
\usepackage[usenames]{color}
\usepackage{euscript}
\usepackage{amssymb}
\usepackage{journals}
\definecolor{grey}{rgb}{0.5,0.6,0.7}

\title[Masses of Fornax UCDs]{Dynamical versus Stellar Masses of Ultracompact
Dwarf Galaxies in the Fornax Cluster\thanks{Based on the archival ESO VLT
data (programme 078.B-0496) available through http://archive.eso.org/
and Hubble Space Telescope archival data (programmes 8090 and 10129)
available through http://hla.stsci.edu/}}
\author[I. Chilingarian et al.]{Igor V. Chilingarian$^{1,2}$\thanks{E-mail:
    igor.chilingarian@astro.unistra.fr, chil@sai.msu.ru}, Steffen Mieske$^{3}$, Michael Hilker$^{4}$ 
    and Leopoldo Infante$^{5}$\\
$^{1}$Centre de Donn\'ees astronomiques de
Strasbourg -- Observatoire de Strasbourg, CNRS UMR~7550, \\ Universit\'e de
Strasbourg, 11 Rue de l'Universit\'e, 67000 Strasbourg, France\\
$^{2}$Sternberg Astronomical Institute, Moscow State University, 13 Universitetski prospect, 119992, Moscow, Russia\\
$^{3}$European Southern Observatory, Alonso de Cordova 3107, Vitacura, Santiago, Chile\\
$^{4}$European Southern Observatory, Karl-Schwarzschild-Strasse 2, 85748 Garching bei M\"unchen, Germany\\
$^{5}$Departamento de Astronom\'ia y Astrof\'isica, Pontificia Universidad
Cat\'olica de Chile, Casilla 306, Santiago 22, Chile
}
\begin{document}

\date{Accepted 2010 Nov 9. Received 2010 Oct 29; in original form 2010
Jul 7}

\pagerange{\pageref{firstpage}--\pageref{lastpage}} \pubyear{2010}

\maketitle

\label{firstpage}

\begin{abstract}
The origin of ultracompact dwarf (UCD) galaxies, compact extragalactic
stellar systems, is still a puzzle for present galaxy formation models. We
present the comprehensive analysis of high resolution multi-object
spectroscopic data for a sample of 24 Fornax cluster UCDs obtained with VLT
FLAMES. It comprises previously published data for 19 objects
\citep{Mieske+08} which we re-analysed, including 13 with available HST
photometric data. Using Virtual Observatory technologies we found archival
HST images for two more UCDs and then determined their structural
properties. For all objects we derived internal velocity dispersions,
stellar population parameters, and stellar mass-to-light ratios $(M/L)_{*}$
by fitting individual simple stellar population (SSP) synthetic spectra
convolved with a Gaussian against the observed spectra using the {\sc
NBursts} full spectral fitting technique. For 14 objects we estimated
dynamical masses suggesting \emph{no dark matter} (DM) in 12 of them and
\emph{no more than 40~per~cent DM mass fraction} in the remaining two, in
contrast to findings for several UCDs in the Virgo cluster. Some Fornax UCDs
even have too high values of $(M/L)_{*}$ estimated using the Kroupa stellar
initial mass function (IMF) resulting in \emph{negative} formally computed
DM mass fractions. The objects with too high $(M/L)_{*}$ ratios compared to
the dynamical ones have relatively short dynamical relaxation timescales,
close to the Hubble time or below. We therefore suggest that their lower
dynamical ratios $(M/L)_{\mbox{dyn}}$ are caused by low-mass star depletion
due to dynamical evolution. Overall, the observed UCD characteristics
suggest at least two formation channels: tidal threshing of nucleated dwarf
galaxies for massive UCDs ($ \simeq 10^8$ $M_{\odot}$), and a classical
scenario of red globular cluster formation for lower-mass UCDs ($ \lesssim
10^7$ $M_{\odot}$).
\end{abstract}

\begin{keywords}
galaxies: dwarf -- galaxies: elliptical and lenticular, cD -- 
galaxies: evolution -- galaxies: stellar content --
galaxies: kinematics and dynamics 
\end{keywords}

\section{Introduction} 

Ultra-compact dwarf galaxies \citep{Hilker+99,Drinkwater+00,PDGJ01}
initially discovered as extragalactic sources unresolved from the
ground-based observations, represent a new class of compact stellar systems
(CSS) observed in the nearby Universe
\citep{MHI04,Hasegan+05,Jones+06,Mieske+07,Mieske+08,MMH08}. At least an
order of magnitude smaller ($10 < R_e < 100$~pc) than M32-like compact
elliptical (cE) galaxies, but still significantly larger than globular
clusters \citep{Drinkwater+03,Jordan+05}, UCDs are best studied in the two
nearby clusters of galaxies: Fornax and Virgo
\citep{Hasegan+05,Jones+06,Evstigneeva+07,Hilker+07,Mieske+08}. Typically,
UCDs have luminosities between $-13.5<M_V<-10.5$~mag and masses about $2
\cdot 10^6<M<10^8 M_{\odot}$ \citep{Mieske+08}.

UCDs, initially defined on a morphological basis, may indeed represent a
heterogeneous class of objects \citep{MHIJ06} of different origins.  The
concepts of UCD formation include: (1) very massive globular clusters having
the same origin as ``normal'' ones \citep{MHI02}, however brightest UCDs
($M_V<-12$~mag) are too bright and frequent to be statistically accounted
for by the Gaussian representation of the globular cluster luminosity
function; (2) stellar superclusters formed in gas-rich mergers of galaxies
\citep{FK02,FK05}; (3) end-products of small-scale primordial density
fluctuations in dense environments \citep{PDGJ01}; (4) tidally stripped
nucleated dEs (dEN's, \citealp{BCD01,Goerdt+08}) or simply dEN's with very
low surface brightness outer components (cf. \citealp{Drinkwater+03}). None
of the concepts is completely ruled out or confirmed yet.

The dynamical M/L ratios of UCDs are on average about twice as large as
those of Galactic globular clusters of comparable metallicity
\citep[e.g.][]{DHK08,Mieske+08,FLGS08,Taylor+10}, although environmental
differences exist (see below). It has been shown that about 1/4 of the M/L
offset with respect to globular clusters is probably due to dynamical
evolution of the latter \citep{KM09}. The remaining offset in M/L indicates
that UCDs may mark the on-set of dark matter domination in small stellar
systems \citep{Gilmore+07,Goerdt+08}, or indeed probe a variation of the IMF
\citep{DHK08,MK08}.
  
\citet{Mieske+08} state a need in observations aimed at the studies of UCDs
stellar populations and age determination in particular, required to explain
the low dynamical $M/L$ ratios in Fornax UCDs compared to those in the Virgo
cluster.  Here we re-analyse the same observational data applying a powerful
full-spectral fitting technique allowing us to obtain simultaneously
internal kinematics and stellar population properties in the largest UCD
sample available until now.  We also improve some data reduction steps,
necessary because of the more stringent requirements for stellar population
analysis regarding sky subtraction. Another motivation for
re-analysing these data was to double-check the velocity dispersion
measurements for certain UCDs. \citet{Mieske+08} used late-type giant stars
from $\omega$~Cen having metallicities of around $-1.0$~dex as stellar
templates. While this is close to the average UCD metallicity, UCDs span a
wider range of metallicities and for metal-rich objects the template
mismatch due to the metallicity difference may affect velocity dispersion
measurements as suggested in \citet{Chilingarian06}.

The paper is organised as follows: in Section~2 we describe spectroscopic
data reduction and analysis, as well as complementary archival HST data used
to extend the sample of UCDs with known structural parameters; the results
of the full spectral fitting are presented in Section~3; comparison of
kinematical and stellar population properties of UCDs with early-type
galaxies and a general discussion about the origin of UCDs is given in
Section~4.

\section{Data: Sources, Reduction, Analysis}

\subsection{Spectral data}

We used the data obtained in the course of the study of CSSs in the Fornax
cluster (program 078.B-0496, P.I.: L.~Infante). The datasets are publicly
available through the ESO Data Archive\footnote{http://archive.eso.org/}. 
The observations were collected at the ESO Very Large Telescope with the
FLAMES/Giraffe spectrograph \citep{Pasquini+02} in the multi-object
``MEDUSA'' mode, using the HR9 setup giving a resolving power
$R\approx17000$ in the wavelength range 5120--5450~\AA\ in the service mode
in 14 observing blocks (OB) between June and December 2007 with the total
integration time of 54\,000~sec. All particular details about observations are
given in \citet{Mieske+08}. We use the distance modulus of the Fornax
cluster $m-M = 31.39$~mag \citep{Freedman+01} corresponding to the distance
19~Mpc and the spatial scale 92~pc~arcsec$^{-1}$.

In order to perform the stellar population analysis we had to improve the
data reduction and sky subtraction with respect to the requirements for the
kinematical analysis presented in \citet{Mieske+08}, therefore we had to
introduce several additional steps before and after standard ESO FLAMES
pipeline data reduction.

The first additional pre-processing step is to apply the Laplacian cosmic
ray cleaning algorithm \citep{vanDokkum01} to the original science frames
and mask all the regions affected by cosmic ray hits in every individual
frame.

The second step, the diffuse light subtraction, is of a great importance for
the success of our study. The simultaneous calibration (SimCal) Th-Ar lamp
inside the spectrograph was switched on during the observations. The average
flux level in the SimCal fibres exceeded those of scientific targets by at
least two orders of magnitude. Therefore, the scattered light from bright arc
lines at a level of 1--2~per~cent severely contaminated the neighbouring
fibres. There is also a smooth diffuse light component in the spectrograph,
having however, only little effects on the flat-fielding.

To account for this contamination, we created a dedicated scattered
light modelling algorithm. Since FLAMES fibres are sparsely placed on a CCD
plane: the median inter-fibre distance is 14.3~pixels while the fibre $FWHM$
is around 4~pixels, therefore it is possible to use inter-fibre gaps
to estimate the scattered light contribution. We use fibre traces derived
from a flat field to identify inter-fibre gaps which normally should not
contain any signal. Then, we assume that all signal detected in the gaps
represent a superposition of scattered light and global diffuse light. In
the beginning we smooth this contribution along dispersion with a boxcar of
4~pix, then we fit it across dispersion at every position along fibre traces
between each pair of SimCal fibres using the 3-rd order smoothing splines. 
We use the spline interpolation instead of smoothing splines in a small
region of the frame containing a parasite signal generated by the
CCD read-out amplifier.

\begin{figure}
\includegraphics[width=\hsize]{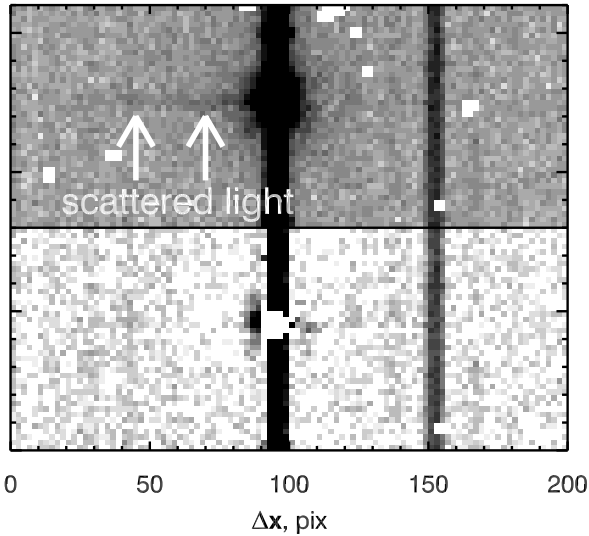}
\caption{A fragment of the science frame before (top) and after (bottom)
diffuse light subtraction. The dispersion is in the vertical direction. The
bright trace in the middle is the SimCal fibre, the one to the right is one
of the brightest scientific targets ($F$-$19$).
\label{figsc}}
\end{figure}

The two-dimensional scattered light model constructed in this fashion is
created and subtracted from every spectral image. In Fig~\ref{figsc} we present
a fragment of a spectral image before and after scattered light subtraction.

Then, the corrected files are used to feed the ESO FLAMES data reduction
pipeline {\sc esorex} to proceed through all steps up-to the linearisation
of extracted spectra.

Then, we apply the third, post-processing step including the adjustment of
the wavelength scale to account for heliocentric radial velocity
corrections, sky subtraction, and combination of spectra for multiple
exposures. All further data analysis is performed on individual
one-dimensional extracted spectra of every object.

We fitted the high-resolution {\sc pegase.hr} \citep{LeBorgne+04} simple
stellar population (SSP) models against the observational data using the
modification of the {\sc NBursts} full spectral fitting technique
\citep{CPSA07,CPSK07}. The original version of {\sc NBursts} cannot be used
directly with our data as we did for FLAMES-LR04 observations
\citep{Chilingarian+08,CCB08}, because the spectral resolution of FLAMES in
the HR9 setup exceeds that of the {\sc pegase.hr} stellar
population models ($R \approx 10000$). Therefore, we degraded the spectral
resolution of the data in order to match that of {\sc pegase.hr} and be able
to provide unbiased estimates of stellar velocity dispersions.

We used the high signal-to-noise twilight spectra obtained in the HR9 setup
of FLAMES available for download from the instrument web-pages. Then, we
fitted the spectra in 5 segments along the wavelength range in every fibre
against the Solar spectrum from the {\sc elodie.3.1} library \citep{PSKLB07}
using the {\sc ppxf} method by \citet{CE04}. The procedure was similar to
the determination of the line-spread-function (LSF) variations described in
\citet{Chilingarian+08} used for low-resolution FLAMES spectra with the
principal difference that now FLAMES-HR9 twilight spectra served as
``templates'' and {\sc elodie.3.1} spectrum as ``data''. The derived
differential spectral line spread is almost constant along the wavelength
direction and does not exhibit any significant fibre-to-fibre variations.
Its shape can be described as a Gauss-Hermite function with $v_0 =
0$~km~s$^{-1}$, $\sigma = 12$~km~s$^{-1}$, $h3 = 0$, $h4 = -0.05$. Then, the
data convolved with this parametrization matched the resolution of {\sc
pegase.hr} models. This procedure is very similar to the degradation of the
{\sc pegase.hr} resolution we performed in all previous studies using the
{\sc NBursts} technique to take into account spectrograph's LSF.

The contamination of UCD spectra by the light of NGC~1399 is an important
source of biases for the stellar population parameter estimates. As it
creates an additive background with very smooth spectral features due to the
high intrinsic velocity dispersion of the galaxy ($\sigma >
200$~km~s$^{-1}$), the metallicity measurements can become underestimated
while velocity dispersion measurements become overestimated \citep[see
Appendix~A3 in][for details]{CPSA07}. These effects should become very
important in the inner region of the cluster.  The sky fibres were placed
far away from the galaxy centre and were hardly a subject to any
contamination by the NGC~1399 halo, therefore we do not expect any
over-subtraction for UCDs located at large projected distances from
NGC~1399.

To account for NGC~1399's halo contamination, we used its spectrum at
the position ($\alpha, \delta)_{J2000} = (54.625042^{\circ},
-35.451722^{\circ}$). We adopted the NGC~1399 centre coordinates from
HyperLeda\footnote{http://leda.univ-lyon1.fr/}: ($\alpha, \delta)_{J2000} =
(54.621208^{\circ}, -35.450667^{\circ}$).

Then we used its parametrized $R$-band light profile from \citet{Dirsch+03}
and projected distances of all our UCDs to scale the NGC~1399 spectrum,
which was then subtracted from UCD spectra. The NGC~1399 spectrum was
smoothed using a $b$-spline with equidistant nodes every 4~\AA. This
suppresses additional noise while subtracting it from UCD spectra and, on
the other hand, keeps enough spectral resolution not to smear absorption
lines in the galaxy spectrum. We estimate the contamination of UCD spectra
by comparing the median fluxes in the scaled NGC~1399 spectrum and original
UCD data. The results (estimated fractions in per cent from the total flux)
are provided in Table~\ref{tabkinstpop}. The strongest contaminated objects
are $F$-$13$, $F$-$17$, and $F$-$11$.

We deduced kinematical and stellar population parameters by fitting the
combined FLAMES-HR9 spectra of UCDs corrected by the contamination of the
NGC~1399 halo using the {\sc NBursts} technique with a grid of simple
stellar population (SSP) models computed using the \citet{KTG93} stellar
initial mass function (IMF). In addition to the best-fitting values of
radial velocity, velocity dispersion, SSP-equivalent age and metallicity
which are obtained in a single minimization loop, for every UCD we
computed a confidence map in the age--metallicity space similar to those
presented in \citep{CCB08}. We notice, that when the {\sc NBursts}
technique is used in the $\chi^2$-mapping mode, i.e. fitting a single SSP
with fixed age and metallicity at every grid node, the algorithm
``degenerates'' into the form totally equivalent to the {\sc ppxf} procedure
by \citet{CE04} with a single template spectrum. Then we interpret the
obtained $\chi^2$ map in terms of stellar populations. In Appendix~A we
present the numerical experiment aimed at quantifying the spectral
information contained in different absorption features present in the
spectral range of our observations. We show that the full spectral fitting
technique in our spectral range using high-resolution {\sc pegase.hr} models
successfully breaks the age--metallicity degeneracy and can be used to
estimate stellar population parameters of old stellar populations from
FLAMES/Giraffe HR09 data.

In principle, UCDs may contain a mix of different stellar
populations so that their spectra will not be well represented by SSP models.
The {\sc NBursts} technique can be used to fit multi-component stellar
population models, however, quite low signal-to-noise ratios of our spectra
did not allow us to do it for every object. It was indeed possible for a
few brightest ones, such as $F$-19, were we attempted to fit two SSPs. In
all cases, the light fractions of young (or intermediate) population were
found to be zero, while old components were identical to the single SSP
cases.

\begin{table}
\caption{Coordinates of 6 UCDs excluded from the final sample of
\citet{Mieske+08}.\label{tabadducdhr9}}
\begin{tabular}{llcc}
id & id$^{*}_{\mathrm{lit}}$ & R.A.($J2000$) & Dec.($J2000$) \\
\hline
$F$-$13$ &             & 03:38:29.16 & -35:27:19.9\\
$F$-$20$ &             & 03:38:56.21 & -35:24:48.9\\
$F$-$28$ & FCOS~2-2106 & 03:38:25.05 & -35:29:25.2\\
$F$-$31$ &             & 03:38:19.78 & -35:23:39.5\\
$F$-$46$ & FCOS~0-2032 & 03:38:30.22 & -35:21:31.3\\
$F$-$62$ &     gc319.1 & 03:38:49.85 & -35:23:36.0\\
\hline
\end{tabular}

\footnotesize{$^*$ identification from \citet{MHI04} for $F$-$28$ and 
$F$-$46$ and from \citet{Bergond+07} for $F$-$62$.}
\end{table}

\begin{figure}
\includegraphics[width=\hsize]{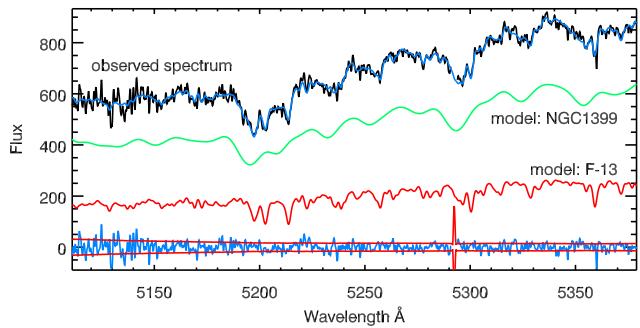}
\caption{Two-component stellar population fitting of the $F$-$13$ spectrum.
The observed spectrum is shown together with its best-fitting template and
its decomposition into two components, NGC~1399 (green) and a UCD (red).
\label{figF13spec}}
\end{figure}

\begin{figure*}
\includegraphics[angle=90,width=\hsize]{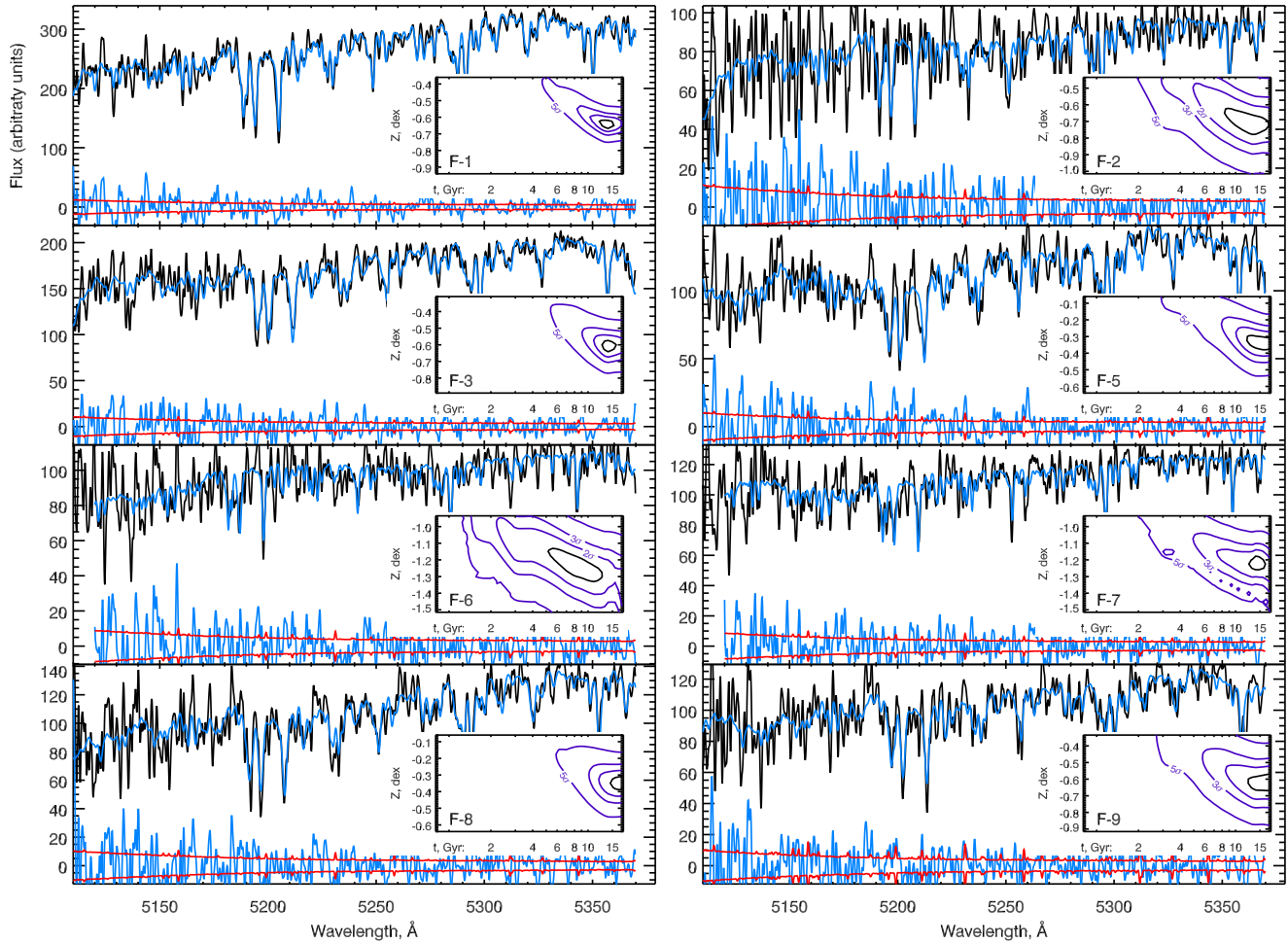}
\caption{FLAMES/Giraffe HR9 spectra, their best-fitting templates (Kroupa 
IMF), fitting residuals and confidence levels of the age and metallicity
determinations (inner panels). Spectra, best-fitting templates, and
fitting residuals were smoothed with a box of 7~pix for clarity. The flux
uncertainties were reduced by a factor of 2.65 correspondingly.
\label{figspec1}}
\end{figure*}

\begin{figure*}
\includegraphics[angle=90,width=\hsize]{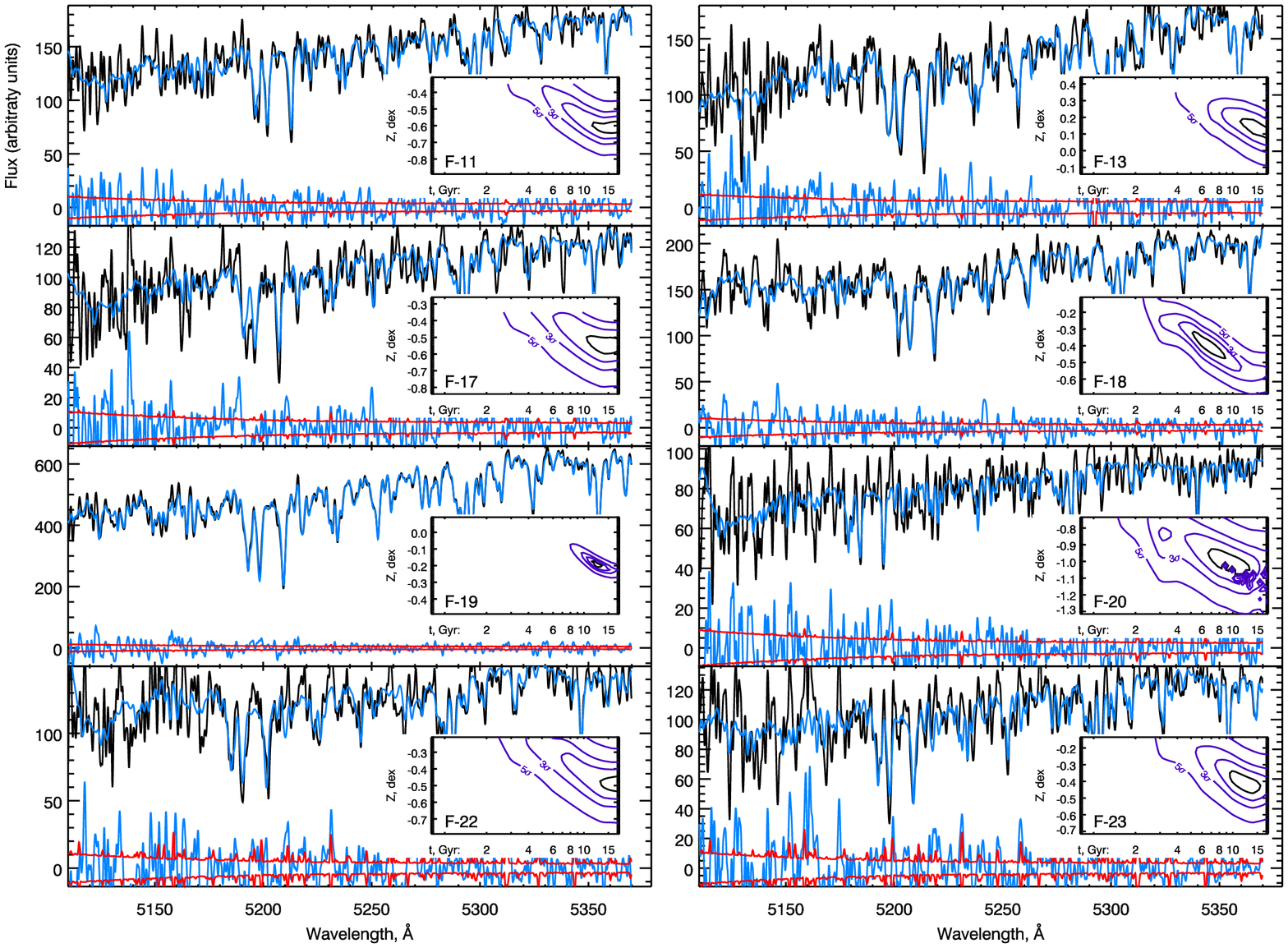}
\caption{Continuation of Fig~\ref{figspec1}.\label{figspec2}}
\end{figure*}

\begin{figure*}
\includegraphics[angle=90,width=\hsize]{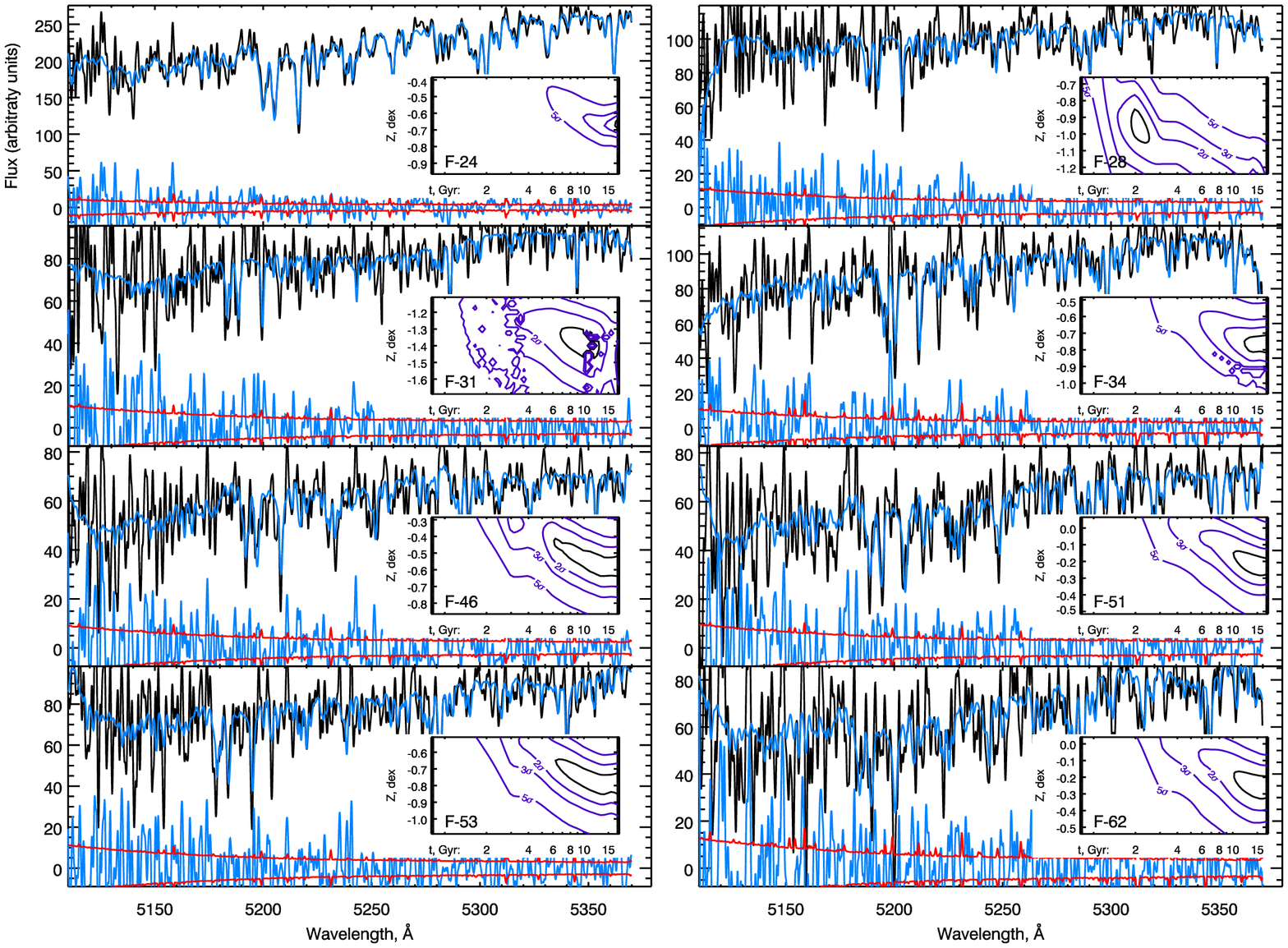}
\caption{Continuation of Fig~\ref{figspec2}.\label{figspec3}}
\end{figure*}

Our final sample includes 19 out of 23 UCDs presented in \citet{Mieske+08}.
Two of the four excluded objects, $F$-$12$ and $F$-$59$ had obvious
artefacts in the sky subtraction due to their very close position to SimCal
fibers, while $F$-$60$ and $F$-$64$ had too low signal-to-noise ratios in
addition to difficulties with the sky subtraction. We are adding 6 objects
to the sample of \citet{Mieske+08} for which the data were obtained, but
they were dropped from the final selection, $F$-$13$, $F$-$20$, $F$-$28$,
$F$-$31$, $F$-$46$, and $F$-$62$. Their coordinates are given in
Table~\ref{tabadducdhr9}. Four of them, $F$-$20$, $F$-$28$, $F$-$31$, and
$F$-$46$, have low metallicities (see next section). None of them but
$F$-$13$ has archived HST data available. For $F$-$13$ it is impossible to
perform surface photometry with reasonable precision, because it is located
0.3~arcmin in projection from the photometric centre of NGC~1399 and
therefore it is projected on the bright part of its spheroid resulting in
heavy light contamination.

The approach we use to correct UCD spectra for the NGC~1399 halo
contamination may introduce biases in case of: (1) highly eccentric or
distorted isophotes of NGC~1399; (2) strong metallicity and/or velocity
dispersion gradients in the NGC~1399 halo. According to \citet{Dirsch+03},
the structure of the NGC~1399 halo is very regular; the velocity dispersion
profile of NGC~1399 flattens out at a level of about
230~km~s$^{-1}$ beyond 15~arcsec \citep{SKGB00} dropping down only
beyond 500~arcsec \citep{Schuberth+10} where its surface brightness is too
low to cause any contamination to the CSS spectra. The optical colour profile
is quite steep \citep{Dirsch+03} which may be indicative of a metallicity
gradient. However, the effects of age and dust may also play a role.

In order to test whether our halo correction results are strongly
affected by radial gradients in NGC~1399, we fitted the uncorrected spectra
of the three heavily contaminated UCDs mentioned above with the model
including two SSPs with different line-of-sight velocity distribution
corresponding to a UCD and the NGC~1399 halo. At first, we left free all
kinematical and stellar population parameters of the two components. The
fitting results for $F$-$13$ are shown in Fig~\ref{figF13spec}. Then we
repeated the test fixing the velocity dispersion of NGC~1399 to
230~km~s$^{-1}$. In both experiments, the UCD kinematical and stellar
population parameters, as well as the recovered UCD spectra were in
excellent agreement. For $F$-$17$ and $F$-$11$ they also precisely matched
the values obtained from the fitting of halo corrected spectra. The
recovered relative mass contributions of two SSP models obtained from the
two-component fitting converted into light contributions are also in a good
agreement with the values estimated from the light profile of NGC~1399.
However, for $F$-$13$, the NGC~1399 light contribution to the total spectrum
recovered from the two-component fitting (73~per~cent) turns to be lower
than that determined from the brightness profile (83~per~cent). This results
in a 0.3~dex lower metallicity (-0.2$\pm$0.09~dex) and 6~km~s$^{-1}$ higher
velocity dispersion (38$\pm$2~km~s$^{-1}$) for the two-component fitting.

From this test we conclude that for UCD spectra weakly and moderately
(up-to 35~per~cent) contaminated by the NGC~1399 halo, the correction
technique based on re-scaling the NGC~1399 spectrum according to its
brightness profile, is consistent with the direct two-component fitting of
spectra. However, the latter approach becomes rather unstable for weakly
contaminated spectra because the low overall signal-to-noise ratio of our
data can not be compensated by high spectral resolution for the NGC~1399
component due to its very high intrinsic velocity dispersion.

\subsection{Imaging data}

Using Virtual Observatory tools, we identified 8 UCDs listed in table~3 of
\citet{Mieske+08} as objects without available HST photometry. For two of
them, $F$-$2$ and $F$-$8$, \emph{F606W} band archival HST WFPC2 $F606W$
band images available through the Hubble Legacy Archive
(HLA)\footnote{http://hla.stsci.edu/} were found and analysed. The data
discovery and access procedure comprised several steps:

\begin{enumerate}

\item The reduced and combined FLAMES dataset was loaded into the {\sc
VO-Paris Euro3D client}\footnote{http://voplus.obspm.fr/$\sim$chil/Euro3D/},
a dedicated Virtual Observatory tool to work with multi-object and 3D
spectorsopic datasets \citep{Chilingarian+08b}. The current version of the
{\sc VO-Paris Euro3D client} includes the native support of the FLAMES FITS
data format.

\item Then, the fibre coordinates were sent to {\sc cds aladin}
\citep{Bonnarel+00} from the {\sc VO-Paris Euro3D client} using the {\sc
plastic}\footnote{http://plastic.sourceforge.net/} application messaging
protocol and then displayed on-top of a DSS2 image of the central part of
the Fornax cluster. The 8 objects reported not to have HST imaging by
\citet{Mieske+08} were selected from the list.

\item Then, a footprint, i.e. a union of all fields of views of all HST
imaging observations in a given area of the sky were requested from the HLA
and displayed using {\sc cds aladin}. Two out of eight objects were
highlighted as those covered by the footprint.

\item Selecting those two objects one by one in the display of {\sc cds
aladin} caused the entries corresponding to the WFPC2 field containing them
to highlight in the HLA Simple Imaging Access query response metadata
tree, i.e. a structured list of HLA datasets. 

\item Then, the corresponding WFPC2 images were retrieved from the HLA and
analysed using the {\sc galfit} software \citep{PHIR02}.

\end{enumerate}

We used the {\sc TinyTim}
software\footnote{http://www.stsci.edu/software/tinytim/} to generate the
WFPC2 PSF at the positions of $F$-$2$ and $F$-$8$ in the
corresponding WFPC2 fields, which was then used during the light
distribution fitting to convolve the input model light profiles.

Both $F$-$2$ and $F$-$8$ were fitted well by single-component
\citet{Sersic68} profiles.  For $F$-$2$ we obtained the following parameters
in the $F606W$ photometric band: $m_{\mathrm{tot}} = 19.98$~mag, $R_{e} =
0.15 \pm 0.01\; \mathrm{arcsec} = 14 \pm 1$~pc, $n=4.9 \pm 0.1$.  For
$F$-$8$, the parameters are: $m_{\mathrm{tot}} = 20.31$ mag, $R_{e} = 0.07
\pm 0.01\; \mathrm{arcsec} = 6 \pm 1$~pc, $n=1.2 \pm 0.1$.

\section{Results and Discussion}

\subsection{Internal dynamics and stellar populations.}

\begin{table*}
\caption{Projected distances, estimated contamination of UCD spectra by the
NGC~1399 light ($C_{\mathrm{h}}$), Internal kinematics, stellar populations, corrected
dynamical and stellar mass-to-light ratios and estimated dark matter content
of 24 UCDs.
\label{tabkinstpop}}
\begin{tabular}{lcccccccccc}
\hline
name & $d_p$ & C$_{\mathrm{h}}$ & $M_V$& $v$ & $\sigma_{\mathrm{obs}}$ & $t$ & $[$Fe/H$]$ & ($M/L$)$_{\mathrm{d,}V}$ & ($M/L$)$_{*V}$ & D.M.\\
     & kpc   & \% & mag & km~s$^{-1}$ & km~s$^{-1}$ & Gyr & dex & ($M/L$)$_{\odot}$ & ($M/L$)$_{\odot}$ & per cent\\
\hline
$   F1$ & 27.5 &  1.7 & -12.19 & 1238.3 $\pm$  0.5 &   22.2 $\pm$  0.6 &   13.8 $\pm$  2.1 &  -0.64 $\pm$  0.02 &  2.8 $\pm$ 0.5  &    3.0 $\pm$  0.2 &  -10 $\pm$ 30  \\
$   F2$ & 33.2 &  3.8 & -11.35 & 1408.5 $\pm$  1.4 &   19.9 $\pm$  1.8 &   12.9 $\pm$  5.4 &  -0.73 $\pm$  0.09 &  3.0 $\pm$ 0.8  &    2.7 $\pm$  0.7 &   10 $\pm$ 50 \\
$   F3$ & 26.3 &  2.7 & -11.71 & 1614.2 $\pm$  1.0 &   33.1 $\pm$  1.1 &   13.9 $\pm$  3.3 &  -0.61 $\pm$  0.03 & $\dots\pm\dots$ &    3.1 $\pm$  0.1 & $\dots\pm\dots$ \\
$   F5$ & 21.5 &  5.7 & -11.73 & 1660.6 $\pm$  0.9 &   25.4 $\pm$  1.0 &   $>$15 $\pm$  3.2 &  -0.34 $\pm$  0.03 &  2.1 $\pm$ 0.3  &    3.8 $\pm$  0.6 &  -80 $\pm$  30  \\
$   F6$ & 14.5 &   12 & -11.07 &  829.0 $\pm$  1.3 &   14.0 $\pm$  2.1 &   11.1 $\pm$  3.2 &  -1.31 $\pm$  0.11 &  1.4 $\pm$ 0.5  &    1.6 $\pm$  0.4 &  -10 $\pm$  60  \\
$   F7$ & 35.2 &  2.6 & -11.13 & 1493.6 $\pm$  0.9 &   12.2 $\pm$  1.6 &   14.8 $\pm$  1.2 &  -1.20 $\pm$  0.04 &  1.7 $\pm$ 0.6  &    2.4 $\pm$  0.1 &  -40 $\pm$  40  \\
$   F8$ & 38.9 &  2.2 & -11.30 & 1395.0 $\pm$  1.2 &   30.2 $\pm$  1.3 &   $>$15 $\pm$  2.8 &  -0.35 $\pm$  0.03 &  4.3 $\pm$ 0.8  &    3.8 $\pm$  0.5 &   10 $\pm$  30  \\
$   F9$ & 58.4 &  1.2 & -11.33 & 1723.8 $\pm$  1.4 &   28.4 $\pm$  1.6 &   $>$15 $\pm$  5.1 &  -0.62 $\pm$  0.04 &  5.7 $\pm$ 1.0  &    3.2 $\pm$  0.8 &   40 $\pm$  40  \\
$  F11$ &  7.7 &   21 & -11.50 & 1686.8 $\pm$  0.8 &   23.7 $\pm$  1.0 &   $>$15 $\pm$  4.8 &  -0.61 $\pm$  0.03 &  1.6 $\pm$ 0.3  &    3.2 $\pm$  0.8 & -100 $\pm$  40  \\
$  F13$ &  1.6 &   83 & -12.08 & 1740.7 $\pm$  1.3 &   32.1 $\pm$  1.5 &   14.0 $\pm$  3.5 &   0.14 $\pm$  0.09 & $\dots\pm\dots$ &    5.3 $\pm$  0.6 & $\dots\pm\dots$ \\
$  F17$ &  7.9 &   26 & -11.27 & 1369.6 $\pm$  1.2 &   28.0 $\pm$  1.4 &   $>$15 $\pm$  4.4 &  -0.55 $\pm$  0.04 &  2.5 $\pm$ 0.5  &    3.3 $\pm$  0.9 &  -30 $\pm$  50  \\
$  F18$ & 36.6 &  1.6 & -11.69 & 2008.9 $\pm$  0.8 &   27.2 $\pm$  0.9 &    6.9 $\pm$  0.8 &  -0.41 $\pm$  0.07 & $\dots\pm\dots$ &    2.0 $\pm$  0.2 & $\dots\pm\dots$ \\
$  F19$ & 45.1 &  0.4 & -13.39 & 1492.0 $\pm$  0.3 &   24.8 $\pm$  0.3 &   12.1 $\pm$  0.6 &  -0.19 $\pm$  0.02 &  4.0 $\pm$ 0.4  &    3.7 $\pm$  0.2 &    8 $\pm$  15  \\
$  F20$ & 32.5 &  3.9 & -11.10 &  671.6 $\pm$  1.1 &   11.9 $\pm$  1.8 &   10.2 $\pm$  3.3 &  -1.02 $\pm$  0.12 & $\dots\pm\dots$ &    2.1 $\pm$  0.5 & $\dots\pm\dots$ \\
$  F22$ & 54.7 &  1.1 & -11.12 & 1039.5 $\pm$  1.3 &   29.1 $\pm$  1.5 &   $>$15 $\pm$  4.4 &  -0.49 $\pm$  0.04 &  2.9 $\pm$ 0.7  &    3.4 $\pm$  0.7 &  -20 $\pm$  50  \\
$  F23$ & 71.3 &  0.8 & -11.66 & 1460.3 $\pm$  1.0 &   17.5 $\pm$  1.2 &   11.9 $\pm$  2.9 &  -0.41 $\pm$  0.10 & $\dots\pm\dots$ &    3.3 $\pm$  0.6 & $\dots\pm\dots$ \\
$  F24$ & 74.7 &  0.4 & -12.27 & 1888.4 $\pm$  0.8 &   29.1 $\pm$  1.0 &   $>$15 $\pm$  2.0 &  -0.67 $\pm$  0.03 &  5.2 $\pm$ 0.7  &    3.1 $\pm$  0.2 &   40 $\pm$  20  \\
$  F28$ & 13.7 &   12 & -10.88 & 1161.0 $\pm$  1.6 &   16.9 $\pm$  2.5 &    2.0 $\pm$  0.4 &  -0.94 $\pm$  0.18 & $\dots\pm\dots$ &    0.7 $\pm$  0.0 & $\dots\pm\dots$ \\
$  F31$ & 21.2 &  7.9 & -10.78 &  927.4 $\pm$  1.4 &   11.0 $\pm$  2.6 &   10.9 $\pm$  3.5 &  -1.39 $\pm$  0.14 & $\dots\pm\dots$ &    1.8 $\pm$  0.5 & $\dots\pm\dots$ \\
$  F34$ & 21.0 &  7.2 & -10.73 & 1621.2 $\pm$  1.1 &   15.3 $\pm$  1.5 &   14.9 $\pm$  4.4 &  -0.77 $\pm$  0.05 &  1.4 $\pm$ 0.4  &    2.9 $\pm$  0.8 & -110 $\pm$  50  \\
$  F46$ & 30.2 &  5.8 & -10.59 & 1424.9 $\pm$  1.5 &   20.0 $\pm$  2.0 &    3.0 $\pm$  0.9 &  -0.32 $\pm$  0.08 & $\dots\pm\dots$ &    3.3 $\pm$  2.0 & $\dots\pm\dots$ \\
$  F51$ & 13.7 &   19 & -10.56 & 1242.9 $\pm$  1.3 &   20.9 $\pm$  1.6 &   $>$15 $\pm$  5.6 &  -0.23 $\pm$  0.11 &  2.8 $\pm$ 0.8  &    4.2 $\pm$  1.2 &  -50 $\pm$  60  \\
$  F53$ & 17.3 &   11 & -10.55 &  660.9 $\pm$  1.3 &   14.5 $\pm$  1.8 &   13.8 $\pm$  5.3 &  -0.80 $\pm$  0.06 &  1.5 $\pm$ 0.5  &    2.9 $\pm$  1.7 & -90 $\pm$  80  \\
$  F62$ & 29.8 &  5.5 & -10.43 &  966.6 $\pm$  1.5 &   19.6 $\pm$  1.8 &   $>$15 $\pm$  6.0 &  -0.26 $\pm$  0.06 & $\dots\pm\dots$ &    4.1 $\pm$  1.3 & $\dots\pm\dots$ \\
\hline
\end{tabular}

\end{table*}

The results of the full spectral fitting are presented in
Table~\ref{tabkinstpop}. We provide radial velocities, internal velocity
dispersions, SSP-equivalent ages and metallicities for all 24 UCDs in our
sample. We obtained the mass-to-light ratios of stellar populations from
{\sc pegase.2} SSP models corresponding to the same ages and metallicities.
We limited the grid of the stellar population models to the maximal
age of 15~Gyr. There is a number of UCDs where the best-fitting age values
corrspond to this limit. Although this value slightly exceeds the presently
adopted age of the Universe, one has to keep in mind that there is a
well-known ``zero-point'' problem in the stellar population models related
to our insufficient knowledge of stellar evolution (see ``the white
paper'' by \citealt{Worthey09} for discussion). Therefore, published ages of
early-type galaxies formally estimated using stellar models often slightly
exceed the Hubble time.

We do not compare our stellar M/L ratios with those obtained using
different stellar population models available in literature because in order
to get consistent results, one has to use the sets of models based on the
same ingredients (e.g. stellar evolutionary tracks, IMFs, binary fractions)
to estimate stellar population parameters from the spectra and M/L ratios
from these parameters. Most ingredients are the same for the {\sc pegase.hr}
and {\sc pegase.2} models. Even though stellar libraries differ, the M/L
ratios of the two sets match in the $V$-band. At the same time, using
different models (e.g. \citealp{BC03a}) for the M/L ratio estimates would
require to use them as well for the determination of stellar population
parameters with the full spectral fitting. However, fitting low resolution
SSP models against our data in a narrow spectral range of the FLAMES
HR09 setup would not allow us to make any sensible estimates of age and
metallicity.

The uncertainties of the stellar population parameters range from 0.03~dex
in $[$Fe/H$]$ and 15~per~cent in age for $F$-$19$ (UCD~3) to $>$0.3~dex in
metallicity and totally uncertain age for the faintest representatives of
our sample such as $F$-$62$. Generally, there is a clear correlation between
the quality of age determination and the overall metallicity of a galaxy,
which can be easily explained as in more metal-rich objects absorption lines
are stronger, thus better constraining the fitting procedure for a given
mean signal-to-noise ratio in the continuum.

Thanks to the high spectral resolution of our data and, therefore a well
sampled line-of-sight velocity distribution in most UCDs, the velocity
dispersion -- metallicity degeneracy \citep[see Section~1.3.1 in][for
details]{Chilingarian06} intrinsic to the full spectral fitting in the pixel
space makes very little effect on the obtained measurements of velocity
dispersion. Hence, even in cases where the metallicity measurements have
very large uncertainties due to low signal-to-noise ratios, the velocity
dispersions still remain well determined.

Our velocity dispersion values sometimes differ significantly from
those obtained by \citet{Mieske+08} (see Fig.~\ref{figsigsig}), therefore their results of
dynamical modelling, i.e. aperture corrections, central and global velocity
dispersions, and, consequently, dynamical mass-to-light ratios have to be
corrected. We repeated the dynamical modelling in exactly the same way as in
\citet{Hilker+07,Mieske+08} using new velocity dispersion estimates and
structural properties from \citet{Mieske+08} for 13 UCDs, and the
photometric properties of $F$-$2$ and $F$-$8$ presented above.

\subsection{Dark matter content and stellar mass function.}

Using dynamical and stellar mass-to-light ratios, we estimate the dark
matter content of 14 UCDs in our sample as $((M/L)_{\mathrm{dyn,}V} -
(M/L)_{*V}) / (M/L)_{\mathrm{dyn,}V}$. The corresponding fractions in per
cent are presented in the last column of Table~\ref{tabkinstpop}. For 9 our
of 14 objects the derived dark matter contents are consistent with zero,
while for $F$-$5$, $F$-$11$, $F$-$34$, and $F$-$53$ the dynamical masses
turn to be lower than the stellar ones resulting in ``negative'' formally
computed dark matter fractions, i.e. $-$100~per~cent corresponding to the
stellar $M/L$ ratio being twice as high as the dynamical one. What is
the reason for derived negative dark matter fractions? We either
underestimate the dynamical mass and/or overestimate the stellar one.

\begin{figure}
\includegraphics[width=\hsize]{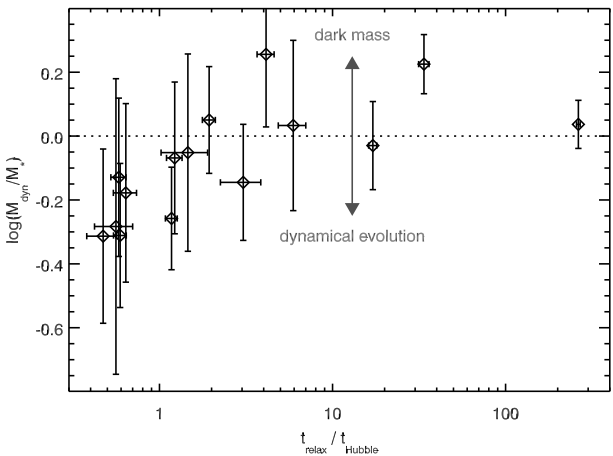}
\caption{Ratio of dynamical to stellar masses vs dynamical relaxation time
at the half-light radius for 14 UCDs with determined dynamical masses. The
dotted line denotes the equality of dynamical and stellar masses for the
assumed stellar mass function. Objects above the line either contain dark
matter, or have either a bottom- or top-heavy IMF; objects below the line
must have their stellar mass functions affected by dynamical evolution.
\label{mltrel}}
\end{figure}

Underestimatation of dynamical masses may originate from imperfect
surface photometry (too small effective radius) or velocity dispersion
measurements biased towards lower values. Although our UCDs have quite small
spatial sizes, most of them are sufficiently well resolved to rule out the
former possibility. The latter option is also hardly possible, because our
data have sufficiently high spectral resolution to reliably measure velocity
dispersions down to 5~km~s$^{-1}$.

Stellar masses may become overestimated if the adopted stellar
mass-to-light ratio is too high. The only parameter which may strongly
affect them at our regime (i.e. old ages, intermediate and low
metallicities) is the shape of the stellar mass function at low masses,
where stars do not contribute much to the total light, but do change
significantly the mass because of strongly non-linear behaviour of the
mass--luminosity relation for main sequence stars. Changing the low mass
slope power law index from 2.3 to 1.3 (i.e. Salpeter to Kroupa) decreases
the mass-to-light ratios of old stellar populations by $\sim$50~per~cent.
Does this suggest IMF variations in the UCDs of our sample?

We have to keep in mind that CSSs observed today might have
experienced dynamical evolution effects on their stellar mass
functions, i.e. the observed mass function may differ from the IMF. It
is known \citep{Spitzer87,BM03, KAS07,KM09} that in globular clusters the
dynamical evolution causes mass segregation, i.e. massive stars are
moving toward the centre while low-mass stars migrate to the cluster
outskirts, where they are tidally stripped during the passages close
to the centre of a host galaxy or through its disc. This creates a
deficit of low mass stars in a cluster, changing the shape of its
integrated stellar mass function. The characteristic timescale of this
process is related to the dynamical relaxation time, which can be
estimated for a CSS \citep{Mieske+08} as $t_{\mathrm{relax}} =
\frac{0.234}{\log M_{\mathrm{dyn}}} \sqrt{M_{\mathrm{dyn}} r_{e}^3 /
0.0045}$~Myr, where $M_{\mathrm{dyn}}$ is in Solar masses and $r_{e}$
is in pc.

In Fig~\ref{mltrel} we present the ratio between dynamical
($M_{\mathrm{dyn}}$) and stellar ($M_{*}$) masses versus the
relaxation time. We see the trend that objects where $M_{*}$ estimates
exceed $M_{\mathrm{dyn}}$ have relaxation times shorter than the
Hubble time. This trend is similar to that presented in Fig~12 of
\citet{Mieske+08}, but in our case the dynamical mass-to-light ratios
are normalised by the stellar ones. It suggests that for some of the
least massive UCDs dynamical evolution is sufficiently advanced to
have experienced preferential loss of low mass stars and a
corresponding flattening of the low-mass stellar mass function, such
that we overestimate stellar masses assuming it to have the Kroupa IMF
shape.

We notice that at the same time very little, if any, dark mass beyond
a canonical Kroupa IMF is required to explain the dynamical $M/L$
ratios of the more massive UCDs investigated here. Since the massive
UCDs described in this study are dynamically un-evolved, the mass
segregation and possible tidal stripping of low-mass stars should not
affect them. Therefore, we can rule out very bottom heavy IMFs like in
\citet{Salpeter55}, which would correspond to 50~per~cent larger stellar
masses and hence imply negative dark matter fraction for all
investigated sources.

There have been various claims in the recent literature regarding a possible
variation or invariance of the IMF at cosmological distances. While
\citet{Cappellari+06,FSB08} argue against a bottom heavy Salpeter type IMF at
high redshift, other gravitational lens results do favour a bottom-heavy
Salpeter IMF \citep{Treu+10,Auger+10,GG10}.  Our result for the investigated
Fornax UCDs is hence in line with the former studies. A caveat is that the
M/L ratios of UCDs appear to vary systematically amongst different
environments \citep{Mieske+08}. The case may indeed be different for
Virgo UCDs which show on average some 40~per~cent higher M/L at comparable
metallicity and age, and which hence appear good candidates for an IMF that
deviates from the Kroupa form \citep{DHK08}.

\subsection{Comparison with literature}

Three objects in our sample, $F$-1 (UCD~2), $F$-19 (UCD~3), and $F$-24
(UCD~4), have published measurements of internal kinematics and stellar
populations \citep{CCB08} obtained using exactly the same data analysis
technique, i.e. the {\sc NBursts} full spectral fitting FLAMES-LR04 spectra
having similar although slightly wider wavelength range but twice lower
spectral resolution. The velocity dispersion measurements agree remarkably
well for all three objects. The metallicity measurements agree within a few
hundredths dex for UCD~3 and UCD~4, however being discrepant by $\sim
0.4$~dex for UCD~2. For this object, the age estimations also differ:
intermediate in \citet{CCB08} and old in our present study. We notice,
however, that the discrepancy of the UCD~2 age and metallicity measurements
in \citet{CCB08} and our present study follow the age--metallicity
degeneracy. Given quite poor data quality and large size of stellar
population confidence levels in \citet{CCB08} we conclude that the
discrepancy between the measurements can be explained by statistical
effects. Age determinations for UCD~3 and UCD~4 agree between the two
studies within 2$\sigma$ of their statistical uncertainties.

It is worth mentioning, that the metallicity estimates for all three UCDs
presented in \citet{MHIJ06} are systematically lower by 0.20$\dots$0.25~dex
compared to our present measurements.

\begin{figure}
\includegraphics[width=\hsize]{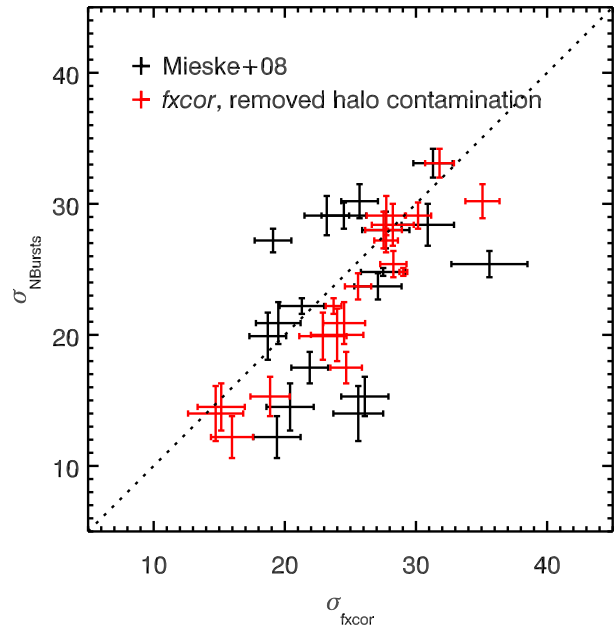}
\caption{Comparison of velocity dispersion estimates from \citet{Mieske+08}
to our present work.\label{figsigsig}}
\end{figure}

In Fig~\ref{figsigsig} (black data points) we present the comparison of
published velocity dispersion measurements for 19 UCDs \citep{Mieske+08}
with those obtained in our study with the {\sc NBursts} full-spectral
fitting technique. Although the general trend agrees, the measurements for
individual objects are often notably discrepant.

The reasons for the discrepancy are:  (1) template mismatch during
cross-correlation due to the metallicity difference between UCDs and
$\omega$~Cen giant stars served as templates; (2) slightly different
wavelength ranges used for the data analysis (inclusion of the Mg$b$ triplet
region in our study); (3) our correction for the contamination of UCD
spectra by the NGC~1399 halo which \citet{Mieske+08} did not apply.

The metallicity difference between the spectra and templates used to
analyse them may lead to biased velocity dispersion measurements (see
Section~1.3.1 in \citealp{Chilingarian06}) at least if the analysis is done
in the pixel space, thus affecting both {\sc NBursts} full spectral fitting
and {\sc fxcor} cross-correlation measurements. The low metallicity of a
template star resulting in shallower absorption lines may be compensated by
decreasing the velocity dispersion, i.e. smearing absorption lines to a lower
degree than it should be in order to match the line depth in the target
spectrum being analysed. Then we would expect to see the correlation between
UCD metallicities and differences of velocity dispersion measurements in
\citet{Mieske+08} and our present study, which we do not detect at a
statistically significant level. This can be explained because the described
degeneracy between metallicity and velocity dispersion mostly affects the
data for targets with velocity dispersions similar to or lower than the
instrumental spectral resolution. In our case, most low-$\sigma$ targets
have low metallicities well corresponding to those of the $\omega$~Cen
template stars. On the other hand, massive metal-rich UCDs with relatively
high velocity dispersions are bright,hence their spectra have good
signal-to-noise ratios reducing the degeneracy effects.

\citet{Mieske+08} had to exclude the Mg$b$ triplet region from their
analysis as it seemed to bias the velocity dispersion measurements obtained
using the cross-correlation technique. However, Mg$b$ is the most prominent
spectral feature in the wavelength range of our spectra, thus containing a
large fraction of spectral information. We performed the {\sc NBursts}
spectral fitting in the wavelength range $\lambda_{\mathrm{restframe}} >
5200$~\AA\ and compared the measurements of velocity dispersion with those
obtained from the fitting in the entire available wavelength range aimed at
checking whether the full spectral fitting technique also suffers from similar
biases. The values turned to be consistent within their uncertainties,
however, being almost half as precise in case of the truncated wavelength
range. Therefore, we conclude that the biases of velocity dispersion
measurements obtained by cross-correlation including the Mg$b$ triplet in
the wavelength range probably originate from the template mismatch when
using stellar spectra as references, which is minimized in our case by
selecting the best-matching SSP from the grid of stellar population models.
As far as these biases affect only measurements of low velocity dispersions
in objects like globular clusters made on relatively high resolution spectra
and do not seem to show up in studies of relatively massive galaxies, we
suppose that this mismatch between the spectra of individual stars and
unresolved stellar populations originates from subtle absorption-line
features and becomes important only at high spectral resolution.

Finally, we repeated the velocity dispersion measurements of
\citet{Mieske+08} using the {\sc fxcor} task in {\sc noao iraf}, but
now applied to the UCD spectra corrected for the contamination of
NGC~1399. Using the same stellar templates of $\omega$ Cen as in
\citet{Mieske+08}, the discrepancy to the results obtained with the
{\sc NBursts} technique has dramatically decreased which is clearly
seen in Fig~\ref{figsigsig} (red data points). We obtain a good
agreement between the two datasets, with only a small residual systematic
offset in the sense that {\sc fxcor} dispersions are slightly larger
than {\sc Nburst} dispersions for low dispersion values. Thus, we
conclude that the contamination of the spectra by the host galaxy halo
is the main reason for biases of the estimated kinematical parameters,
and it has to be taken into account in all studies addressing the
internal dynamics of compact stellar systems.

\subsection{Faber--Jackson and metallicity--luminosity relations.}

\begin{figure}
\includegraphics[width=\hsize]{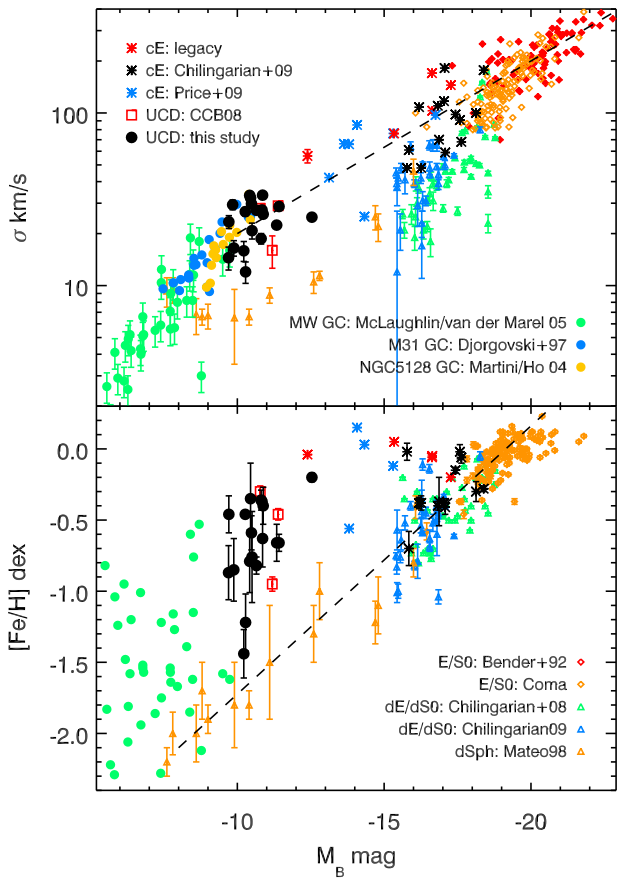}
\caption{Faber--Jackson (top) and metallicity--luminosity (bottom) relations
for early type galaxies and compact stellar systems. The data sources are
given in the text. The dashed line in the upper panel denotes the 
Faber--Jackson relation ($L \propto \sigma^4$) defined for massive 
early-type galaxies.\label{figfjrmet}}
\end{figure}

In Fig.~\ref{figfjrmet} we present the relation between absolute magnitudes
and measured velocity dispersions (top, \citealp{FJ76}) and metallicities
(bottom) for early-type galaxies and CSSs.  The giant
galaxies are represented by 140 early-type galaxies in the Coma cluster from
SDSS DR7 \citep{SDSS_DR7} obtained using the {\sc NBursts} technique and
velocity dispersion measurements from \citet{BBF92}. The data for dE/dS0s
are from \citet{Chilingarian+08} and \citet{Chilingarian09} (global ``main
galactic body'' values). The local group dwarf spheroidals from
\citet{Mateo98} are at the low-luminosity end of the diagrams. The data for
cE galaxies including ``legacy'' objects are from \citet{Chilingarian+09},
transitional cE/UCD galaxies are from \citet{CM08} and \citet{Price+09}, and
three additional UCDs are from \citet{CCB08}. The globular cluster data come 
from \citet{MvdM05,Djorgovski+97,MH04} for Milky Way, M~31, and NGC~5128 
clusters respectively.

In the upper panel of Fig.~\ref{figfjrmet}, we see that all new kinematical
measurements of UCDs follow the faint magnitude extrapolation of the
Faber--Jackson relation ($L \propto \sigma^4$) defined initially for giant
early-type galaxies and shown as a dashed line in the upper panel of
Fig.~\ref{figfjrmet}. Most cEs and transitional cE/UCD galaxies reside on
it. At the same time, dEs, dS0s and the dark matter dominated dSph galaxies
deviate from the Faber-Jackson relation exhibiting lower velocity dispersions for the same
luminosity.

Globular clusters form a different sequence on this diagram having a
steeper slope ($L \propto \sigma^2$) than the Faber--Jackson relation.
UCDs of our sample populate the low-luminosity extension of the
Faber--Jackson sequence down to luminosities where it joins with that of
globular clusters creating a smooth transition.

On the metallicity--luminosity diagram the situation is notably
different: the $Z - L$ relation denoted by a dashed line in the lower
panel of Fig.~\ref{figfjrmet}, is formed by dSph, dE/dS0, and giant
early-type galaxies, whereas the compact galaxies are generally more
metal-rich and do not show such a correlation between their
metallicities and luminosities filling the entire region above the $Z
- L$ relation up-to the solar metallicity. The offset metallicity with
respect to the extended dwarf spheroidals is consistent with a more
efficient self-enrichment, and/or them being formed from already
pre-enriched material in the course of wet galaxy mergers.

\subsection{Origin of UCD galaxies}

We conclude this paper with a brief discussion on the origin of
UCDs. One can break down the currently discussed formation channels
(see Introduction) into two concepts:
\begin{enumerate}
\item UCDs are tidally stripped remnants of more extended galaxies, hence
of galaxian origin. As such they would also trace the tidal disruption
of low-mass dwarf galaxies.
\item UCDs are massive star clusters -- or mergers thereof -- whose
formation is closely linked to the formation of the bulk of the
globular cluster population.
\end{enumerate}

Regarding the first concept, \citet{Chilingarian+09} demonstrated that a
Milky Way sized disc becomes tidally stripped on a timescale of 0.5--1~Gyr
by an intermediate mass cD galaxy similar to the Virgo cluster M~87. Since
dwarf galaxies usually have lower stellar disc densities, the threshing
should be even more efficient and quick. A single passage near the cD galaxy
should be sufficient, suggesting a characteristic time of the process to be
about 200~Myr. This explains why we do not observe many dEN galaxies during
the process of tidal stripping, this phase is short and therefore it is
statistically unlikely to observe it.

The discovery of the first transitional cE/UCD galaxy M59cO \citep{CM08} and
a population of galaxies in the Coma cluster \citep{Price+09} completely
filling the mass gap between cEs and bright UCDs, gives additional support
to the common evolutionary path followed by members of these two classes.
However, classical cE galaxies can hardly be formed as tidal superclusters
in galaxy mergers because (1) they have very significant masses exceeding
$10^9~M_{\odot}$ which are quite problematic to assemble; (2) all nearby cEs
rotate and contain either confirmed central supermassive black holes
\citep{DBM08} or central bumps in the velocity dispersion distribution
probably caused by them \citep{CB10}.

Thus, if we accept (massive) UCDs and cEs to share the same
evolutionary scenario, they should have formed by the tidal threshing
of more massive progenitors. The clearest candidates for such tidally
stripped nuclei are the two respective most massive UCDs in the Fornax
and Virgo cluster (e.g. \citealp{Evstigneeva+07,Hilker+07}),
which have extended low-surface brightness envelopes. They have masses
$M \simeq 10^8 M_{\odot}$ and clearly stand out among all other UCDs
in terms of their sizes and luminosities, see also the location in the
M$- \sigma$ plane in Fig.~\ref{figfjrmet} for the most massive Fornax UCD.

In contrast, the bulk of fainter, lower mass UCDs show a smooth luminosity
and size transition towards the regime of ordinary globular clusters, and
their number counts are typically well accounted for by an extrapolation of
the globular cluster luminosity function (e.g. \citealp{MHI04}). This is
consistent with a scenario where the bulk of these UCDs (in particular the
metal-rich ones) formed when the majority of the red GCs and the spheroids
of their host galaxies were created (either via a monolithic collapse or
multiple wet mergers).

\section*{Acknowledgments}

IC acknowledges the ESO Visiting Scientist programme. This study is based on
observations made with the European Southern Observatory Very Large
Telescope; partially based on observations made with the NASA/ESA Hubble
Space Telescope, and obtained from the Hubble Legacy Archive, which is a
collaboration between the Space Telescope Science Institute (STScI/NASA),
the Space Telescope European Coordinating Facility (ST-ECF/ESA) and the
Canadian Astronomy Data Centre (CADC/NRC/CSA). This research has made use of
{\sc SAOImage ds9}, {\sc noao iraf}, and {\sc cds aladin} software tools and
packages.

\bibliographystyle{mn2e}
\bibliography{UCD_Fornax_HR}

\appendix

\section{Age and metallicity sensitive spectral information in the
FLAMES/Giraffe HR09 setup}

\begin{figure}
\includegraphics[width=\hsize]{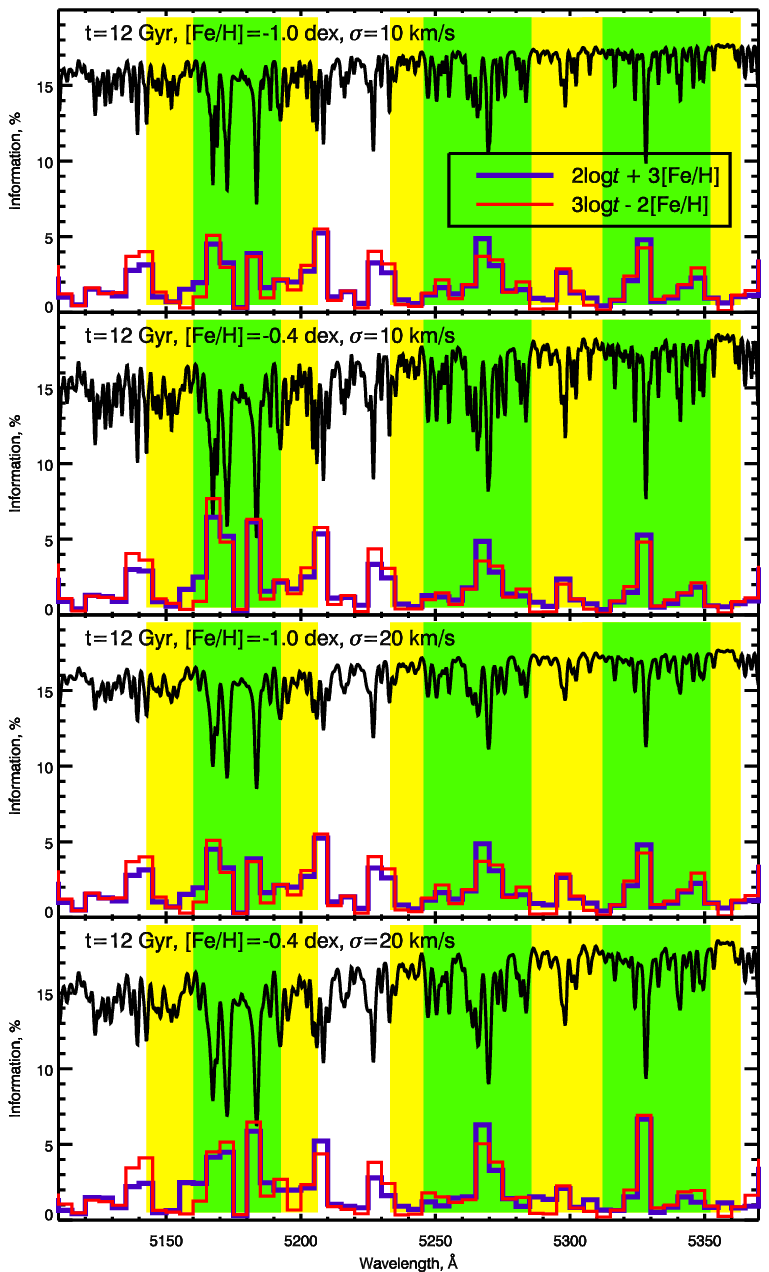}
\caption{Spectral distribution of the stellar population sensitive
information for four stellar populations in 5~\AA-wide bins in the
FLAMES/Giraffe HR09 setup. Corresponding age, metallicity, and velocity
dispersion are indicated in every panel. The studied SSPs are shown in
black; blue and red histograms display the relative importance of spectral
bins (in per cent) for the determination of $\eta$ and $\theta$ stellar
population properties respectively. Green bars denote the bands defining the
following Lick indices: Mg$b$, Fe$_{5270}$, and Fe$_{5335}$ with
yellow side-bars corresponding to the corresponding pseudo-continuum
definition regions. \label{figspecinfo}}
\end{figure}

Using the technique described in \citet{Chilingarian09}, we quantified the
age and metallicity sensitive information in the absorption-line spectra at
intermediate spectral resolution $R=10000$ in the wavelength range
corresponding to the HR09 setup of FLAMES/Giraffe for a single SSP fitting.
We explored the relative sensitivity of different spectral features to the
stellar population parameters defined in the rotated age--metallicity
coordinate system $\eta - \theta$ \citep{Chilingarian+08,Chilingarian09}
with the $\eta$ axis parallel to the age--metallicity degeneracy
\citep{Worthey94}:
\begin{eqnarray}
    \eta = (3 Z + 2 \log_{10} t) / \sqrt{13}; \nonumber\\
    \theta = (-2 Z + 3 \log_{10} t) / \sqrt{13}
\label{coordeq}
\end{eqnarray}

\noindent
The contribution of every pixel at every wavelength to the total $\chi^2$
depends on many parameters. In order to estimate the fitting procedure
sensitivity $S(\lambda, p_0, \dots, p_n)$ to a given parameter $p_i$ at a
given wavelength $\lambda$, we should compute the corresponding partial
derivative of the template grid removing the global continuum shape, i.e.
accounting for the multiplicative continuum variations. Hence, the
response to a stellar population parameter $\eta$ (and similarly,
$\theta$) at a given point $(t_0, Z_0, \sigma_0)$ of the parameter space is
expressed as:
\begin{align}
S(\lambda, \eta)&|_{(t_0, Z_0, \sigma_0)} = \nonumber\\
 & \frac{\partial}{\partial \eta} \chi^2\{P_{1p}(\lambda, \eta, \theta)|_{(t_0, Z_0)}
    (T(\lambda, t, Z) \otimes \mathcal{L}(\sigma_0)) \},
\label{senseq}
\end{align}

\noindent where $\mathcal{L}$ is the LOSVD, $T(\lambda, t, Z)$ is the flux
of a given SSP template spectrum characterised by its age and metallicity at
a given wavelength, $P_{1p}$ is the $p$-th order multiplicative Legendre
polynomial. The meaning of this equation is that the squared fitting
residuals of a spectrum $(\eta_0 + \Delta \eta, \theta_0)$ against the model
$(\eta_0, \theta_0)$ at every pixel would correspond to its contribution to
the overall $\chi^2$ when varying $\eta$.

In practice, we compute these derivatives numerically using a ``one-side''
approach, i.e. computing the function values at a point of interest and at a
nearby point slightly offset from it on a given coordinate.
For our test we chose two SSP models with the age $t=12$~Gyr and
metallicities $-1.0$ and $-0.4$~dex correspondingly. Then we convolved them
with Gaussian kernels corresponding to the internal velocity dispersions of
10 and 20~km~s$^{-1}$ and, thus, obtained four model spectra or ``reference
SSPs''. Then for every of them we varied the $\eta$ and $\theta$
parameters corresponding to their ages and metallicities (see
Eq.~\ref{coordeq}) by 0.05~dex, which corresponds to (12.8~Gyr, $-0.96$~dex)
and (13.2~Gyr, $-1.03$~dex) for the metal-poor models and to (12.8~Gyr,
$-0.36$~dex) and (13.2~Gyr, $-0.43$~dex) for the metal-rich ones. Later, we
fitted these models against their ``reference SSPs'' using the {\sc ppxf}
procedure \citep{CE04} with the 10th order multiplicative polynomial
continuum. The fitting residuals obtained by this procedure correspond to
the quantity defined in Eq.~\ref{senseq}.

We have co-added the information on $\eta$-, and $\theta$-sensitivity in
5~\AA\ bins in the wavelength range between 5100 and 5380~\AA\ and
normalised it by the total value of non-reduced $\chi^2$ thus obtaining the
relative importance of every 5~\AA-wide bin to the determination of the
stellar population parameters. The results for our four ``reference SSPs''
are shown in Fig~\ref{figspecinfo}. One should not directly compare blue and
red curves, because they represent the normalised quantities. The absolute
values of $S(\lambda, \eta)$ as defined by Eq.~\ref{senseq} are 2--2.6
times higher than $S(\lambda, \theta)$ explaining the elongated shapes of
the 1-$\sigma$ uncertainty ellipses in the age--metallicity space.

The age and metallicity response cannot be determined directly using this
approach, because these two parameters are degenerated. Then, the formally
computed sensitivity will become overestimated.

We see that $\eta$-sensitive information (the one allowing us to break the
age--metallicity degeneracy) and $\theta$-sensitive information represented
by red and blue histograms in Fig.~\ref{figspecinfo} respectively are
distributed slightly differently along the wavelength range. In general, the
blue and red histograms are much more similar to each other than in the case
of lower spectral resolution as presented in fig.~2 of
\citet{Chilingarian09}. However, we see in Fig.~\ref{figspecinfo} that at
high spectral resolution the sensitivity across the age--metallicity
degeneracy (blue histogram) is stronger correlated with equivalent widths
of metal absorption lines than that along the degeneracy (red histogram)
which is more bound to the Mg$b$ triplet and fainter absorptions around
5140~\AA\ and 5230~\AA\ but not to the strong iron multiplet defining the
Fe$_{5270}$ Lick index.

The main conclusion we draw from this test is that \emph{when we use the
full spectral fitting at a sufficiently high spectral resolution, then even
in the narrow wavelength range of the FLAMES/Giraffe HR09 setup including no
``traditional'' age indicators such as H$\beta$, we are able to constrain
stellar population age with a high level of accuracy.}

The whole idea of the stellar population determination using the full
spectral fitting is related to the difference of age and metallicity
response functions of different pixels in a spectrum. These are related to
the flux response of a stellar spectrum at a given wavelength to effective
temperature, surface gravity, and metallicity, because we use the stellar
spectra as principal ingredients to build stellar population models. Even if
we deal with the spectral region containing mostly lines of metals, which
become deeper when the stellar population metallicity is increasing or the
population is becoming older, the exact behaviour of individual lines (or
groups of lines) will be slightly different which is again connected to the
absorption line properties in stellar spectra. For example, the behaviour of
the Mg$b$ triplet equivalent width as a function of $T_{\mbox{eff}}$ is not
the same as that of the Fe$_{5270}$ multiplet, while the metallicity
dependence is rather similar. These differences for different sometimes very
faint spectral lines and molecular bands are co-added at every pixel
included in the fitting and at the end bring enough contribution to the
$\chi^2$ in order to determine the age of the stellar population from the
fitting procedure. However, if the intrinsic velocity dispersion of the
stellar system being studied is high (or, alternatively, the spectral
resolution is low), then the age and metallicity sensitive information in
faint absorption line features is rapidly decreasing, hence increasing the
relative importance of the most prominent spectral features as those used to
define the Lick system. Therefore, while we can successfully determine
stellar population properties of UCDs from our FLAMES/Giraffe HR09 data, it
would be virtually impossible to do so for giant early-type galaxies,
because of their high intrinsic velocity dispersions.

\label{lastpage}

\end{document}